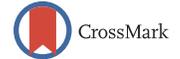

# Research Article
# Traffic Analysis for Storage Finding in Video on Demand System

Soumen Kanrar

Department of Computer Science, Vidyasagar University, Midnapore, 721102 West Bengal, India

## Abstract

**Background and Objective:** The literature survey typically predicated sharp growth for IP-based video traffic i.e., 30% or more annually. For the Internet TV in mobile networks, video traffic growth rate is expected to rise 80% or more. These high growth rates of video traffic will account for a large portion of the bandwidth. The performance of video-on-demand system during real-time data streaming greatly depends on the session oriented data-storage finding in the mass scale distributed storage architecture. At the storage end, data is broken up into manageable chunks of data packets, which could be smoothly, deliver over the Internet. The objective of this study was to present the necessity of traffic control and traffic analysis methodology in the video on demand system to minimize the hop count for finding exact media storage to retrieve video chunk data. **Methodology:** Multiple inbound and outbound connections virtually appear a single connection to the user. The session based storage finding mechanism emphasis inbounds paths from distributed storage. The minimum hop counts for storage finding effectively reduce the search cost in video on demand system. For the enhancement of overall system performance, Zipf approximation was used particularly for the outbound traffic requests from the user end. The LRFU (least recently frequently used) mechanism was implemented on the 'web cache' at storage node with a considerable cache hit ratio. The inbound traffic flows for BGP using the AS_PATH metric to avoid loops. If the route paths are not locally organized, then the route path uses the AS_PATH attribute to ties between the weight and local preference attributes. The attributes are used to select a particular path that controls inbound traffic. **Results:** This study presents novel solutions regarding the existing issues on the video traffic flow. Three stages of simulation have been observed according to the aforesaid methodology. In the first stage of simulation, traffic analysis determined the VOD system. In the second stage of simulation, it has been considered distributed database storages. In third stage number of phases considered to make complete simulations. **Conclusion:** Traffic control and traffic analysis methodology for video on demand system minimizes the hop count during exact storage search.

Competing Interest: The author has declared that no competing interest exists.

Data Availability: All relevant data are within the paper and its supporting information files.

## INTRODUCTION

The network traffic forecasts, such as cisco visual network index[1], predict strong growth rates for video traffic. Typical predicated annual growth rates increase 30% or higher for wire-line IP-based video services and in the case of mobile networks and Internet TV likely to 90%. These massive growth rates, video traffic will occupy the large portion of bandwidth in communication networks. According to Cisco Inc., prediction video will contribute close to two-thirds of the mobile network traffic, at the end of this decade. So the network designers and technologist are required to better understanding of video traffic, in order to account for the video traffic characteristics in designing and evaluating communication services for this emerging type of network traffic. Asynchronous arrival of request, users wish to watch videos from beginning as well as willing to avail interactive services like jump, rewind, fast-forward and move backward with some frame numbers etc. The VOD system makes it more challenging to design and deployment. A number of researches have already been done on that direction, particularly in the area of architectural design in Video on demand systems. This work addresses some of the novel solutions for the existing challenging issue in the domain of traffic flow. Traffic flow analysis for distributed storage in a course of an interactive session is effectively elevated, to reduce the packet loss and chunk video objects finding cost. Distributed storage architecture enhances, smooth video streaming. The robust traffic flow in the video on demand system reduces the extra load at intermediate nodes in VOD system. The traffic analytics in large scale video surveillance usually select features like on-demand watching of segmented scenes and required to take lists of decision on the fly[2]. It requires how fast the VOD system can retrieve the data and stream that directly to the end user. The rapid modification in the area of computer hardware was carried a new demand to adopt updated methodologies for the massive requests. The massive amount of requests was generated for partial video clips or for complete video that was desired to be delivered within a very short interval of time. The massive number of the submitted requests at the web cache was efficiently approximated by Zipf like distribution. The Zipf approximation approaches were used and addressed in the following references[2-7]. The performance of the system highly depends on the response from the web cache[8,9]. The enhancement of the storage portal depends on the efficient handling of the cache memory[9-11]. Researchers have proposed various types of concepts over the last decade to handle the cache memory for distributed multi-tier computing. On demand 'audio/video' user grows exponentially due to the growth of the Internet and smart mobile phone. It has been observed that the major traffic load experience at the 'audio/video' portal for some of the sessions. Tian *et al.*[10], proposes multi queue replacement polices at second level buffer cache that can handle the stream of requests. Muntz and Honeyman[12] and Ari *et al.*[13] have observed that due to lack of proper handling polices, web cache experiences a poor hit ratio. Recent trend being observed as the indexing of cache pages in the real time. Zhou *et al.*[14] and Kanrar and Mandal[15] have focused on changing a fresh reference pattern on the cache to improve the smooth video streaming. Video-on-demand applications are becoming very popular and important area for scalable video distribution in academic, commercial environments and surveillance sector, respectively. This works address some of the direct solutions to the traffic analysis and control in video on demand system to enhance the search paths for finding the archived storage in video on demand system.

This study elaborately presents the necessity of traffic control and traffic analysis methodology in the video on demand system to minimize the hop count for finding exact media storage to retrieve video chunk data.

## MATERIALS AND METHODS

**Process flow mechanism:** The class of 'video on demand' users have submitted massive amount of requests over the Internet to access the required video stream (full portion) or partially as a part of its chunk video. The user requests were generally forwarded through the HTTP or HTTPs protocols. The submitted requests were simultaneously processed at the web cache container. The viewer's browser sends the request to a web site. This submitted request in turn generates a basic request to domain name system (DNS) for IP address of the required video content web site. The DNS returns the corresponding IP address of the load balancer for that required video container site. The browser was forwarded the request to the corresponding load balancer. In turn, the load balancer forwards the request to the web server. If the requested video chunk was presented in its web cache memory, then web cache sends the content directly to the browser. The HTTP or HTTPs was requested that web server, where the objects were retrieved from the web cache, without going to the original server. This was referring as cache hit. In a case, if the web cache does not have the requested size of content, completely or partially of it, then it hands over the request to the application web server. This was referring the cache miss. The application web server smoothly distributes its tasks among the multiple data servers inside the storage. It was processed as an individual task to search the full content of the video or part of the video i.e., the chunk of the video stream[16,17]. The complete portion of video or partially the chunk of a video sends back to web cache[13,18]. The web cache sends the video completely or partially on demand basis to the viewer and stores copy of that in cache memory[19,20]. The flow mechanism related to video-retrieves is presented in the Fig. 1. Generally, viewers send the request simultaneously via the Internet. The Fig. 1, presents the access video streams from a particular storage site. Initially, all requests in the form of HTTP or HTTPs requests were forwarded to web cache. If the web cache that has the content of the video stream, then forwarded the content to the viewer directly otherwise the request was forwarded further.

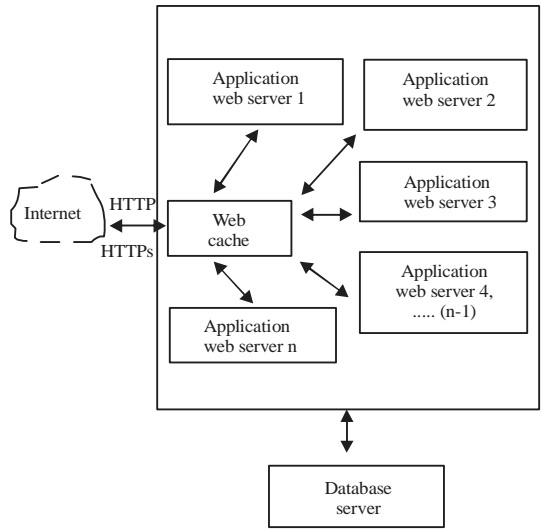

Fig. 1: Architectural view of the integrated storage

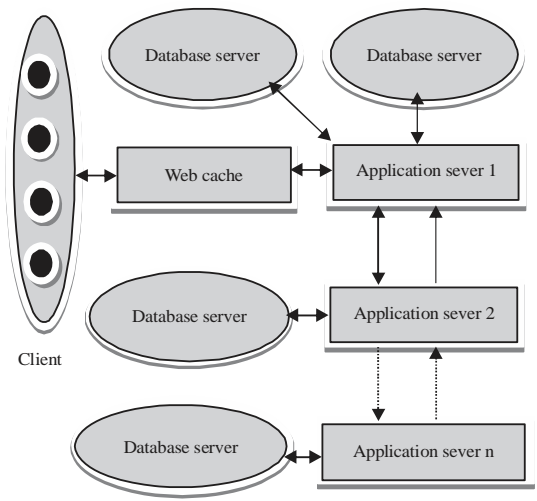

Fig. 2: Distributed orientation of database servers

In case, if the web cache memory does not possess the required content, then the request was forwarded to Application server from the list of application servers[19,20]. The application web server was selected from the least loaded application server. The application web server searches the required video stream from a set of the well-connected distributed database server according to the Fig. 2.

**Inbound traffic mechanism:** Real time multimedia data streaming applications over the Internet are classified into two major forms, likes live and on-demand. Video conferencing, it was the live broadcasting of important events, which relates to the popular applications in the area of live streaming. Whereas, video on demand (VOD) was more suitable for some other suitable applications like on-demand distance learning and pull based content distribution such as on-demand watching of media using internet protocol television etc. The VOD system provides the user with extra additional exibility in terms of independence of watching time, location and device. Thus a lot of research has been dedicated in that direction. The incoming viewers request can be better approximated by the Zipf like distribution[2]. For streaming, three major components were involved, source of media, overlay network architecture and various client requirements. Source of media may be centralized or distributed. In a centralized architecture, clients directly connect to a single source server[21]. It was successful still the web cache has enough resource capacity to handle the incoming requests. Nevertheless, the situation worsens when the demand for media increases beyond the web cache's capacity and the viewer regularly used simultaneous interactive operations[22,23]. The quality of service (QoS) like scalability, robustness and fault tolerance suffers in client-server architecture. The distribution of viewer requests in generally follows a typical Zipf-like distribution, where the relative probability of a request for i (the most popular page) is proportional to $1/i^\alpha$, with a positive real number ($0<\alpha\leq1$). The viewer's request distribution rarely follows the strict Zipf law (for which $\alpha=1$)[24]. For Zipf-like distribution, the cumulative probability that one of the kth popular video was accessed (i.e., the probability of a popular video request) is given asymptotically with the Eq. 1, with real value k:

$$\Psi(k) = \sum_{i=1}^{K} \frac{\delta}{i^\alpha} \approx \delta \frac{k^{1-\alpha}}{(1-\alpha)} \qquad (1)$$

where, $\delta$ is the real value depends on $\alpha$ and N such that $\delta \approx \frac{(1-\alpha)}{N^{(1-\alpha)}}$ and N is the total number of videos present in the system. The asymptotic function $\Psi(k)$ is used to approximate (k/N). It is expressed in the form of Eq. 2:

$$\Psi(k) = (k/N)^{(1-\alpha)} \qquad (2)$$

Since, (k/N) <1 so for all the meaningful values of k and a larger value of $\alpha$ which was increased the value of $\Psi(k)$. Basically, the major viewer requests are concentrated on some popular videos[25]. Based on this phenomenon, the estimated probability of the viewer requests for an unpopular video in the VOD system with the total N number of videos present in the system and out of that, k numbers of videos are popular ones can be presented with the Eq. 3:

$$p_{unp} = 1 - (k/N)^{(1-\alpha)} \qquad (3)$$

In a large-scale video on demand system, bandwidth optimization largely depends on the minimum hop counts during the high demand stage for a specific interval of time or some of the days in a week. Efficiency of the video on demand system practically depends on the effectively handling of 'least recently frequently used' (LRFU) cache memory handling mechanisms at web cache[26,18]. The LRFU mechanism was composed of the two dynamically growing cache memory blocks, one was 'least recently used' (LRU) and another was 'least frequently used' (LFU) cache memory. The video stream retrieves from the storage server depends upon the cache replacement polices. The replacement policy directly depends upon the cache hit ratio and cache miss ratio at real time. As the cache size was limited with compare to the auxiliary memory size and it was only the fraction of the auxiliary memory. The cache memory at the web cache, maintains two blocks one for the least recently used (LRU) and other least frequently used (LFU). The 'least recently frequency used' (LRFU) policy was used to stored pages into the cache memory. Since, the size of the cache was limited. Hence, the cache was used the exponential smoothing parameter that dynamically replaces the page with the smallest hit count from the LRU block. The viewer request makes increment the hit counts for the listed pages at web cache. The implication of LRFU policy and the impact of that policy for the limited cache size at web cache for huge requests in a very small interval of time were great. LRFU policy was combined effect of LRU and LFU. Initially, LRFU policy assigns a value Hit_count (x) = 0 to every page 'x' and that value be updated every instance of time.

It is logically expressed as:

$$Hit\_count = \begin{cases} If\ \{page\ is\ (x)\ referenced\ at\ time\ \Delta t\} \\ \qquad then\ 1 + 2^{-\lambda} \times \left(Hit\_count\right)\} \\ else, \\ 2^{-\lambda} \times \left(Hit\_count\right) \end{cases}$$

The parameter $\lambda$ is an integer variable that controls the behavior of the viewer request flow. The performance of page replacement depends upon the value of $\lambda \in (0,1)$. The performance of the video on demand system purely depends on the efficient cache replacement at the web cache[27]. The viewer request pattern to the web cache was approximately presented by the Zif's probability distribution[8] and it was expressed in Eq. 1 and 2. Bandwidth optimization, for real-time traffic flow through the web cache server was a challenging issue in the high-speed network. The aggregate bandwidth requirement for video file was efficiently approximated during on demand page replacement at the web cache server for any types of the service request. The self-control traffic rate at the web cache memory was optimized based on the relative data present at the web cache server or require to import a particular video file from the distributed storage. The smoothness of traffic transmission was highly depended on the self-configuration of web cache memory at web cache server and page replacement mechanism. The web cache memory was finite, in size. So the enhancement of the system performance depends upon the self-optimization of the traffic rate. The effect of the Zipf distribution parameters for the incoming viewer requests as well as the effects of those parameters with the request was used to measure the bandwidth requirement for the desire video data file. The bandwidth optimization corresponds to the traffic flow was depended on the size of the video file and the active duration of stream data transmission. If the hit misses occurs for the ith file or page according to Eq. 1 and $b_i$ is the required amount of bandwidth to import the ith file from the storage, then:

$$Bandwidth_{demand\ (i)} = b_i \qquad (4)$$

where, $b_i = (s_i, t_i)$ in ideal condition and the variable $s_i$ is the size of the ith requested video file. The variable $t_i$ presents the time occupancy of that channel for data transmission.

The aggregate bandwidth for on demand requested file is approximated by the Eq. 4:

$$\sum_{i=1}^{\infty} H_{demand(i)} \cdot (s_i \cdot t_i) \begin{bmatrix} = \sum_{i=1}^{\infty} \sum_{j=1}^{C} p_N(j)(s_i \cdot t_i) \\ = \sum_{i=1}^{\infty} \{\sum_{j=1}^{C} p_N(j)\}(s_i \cdot t_i) \approx k \sum_{i=1}^{\infty} \alpha \cdot (C^{1-\alpha}) \cdot (s_i \cdot t_i) \end{bmatrix} \quad (5)$$

Since, $\quad p_N(i) \approx \dfrac{1}{i^a} \cdot \dfrac{1}{\displaystyle\sum_{i=1}^{N} \dfrac{1}{i^a}}$

According to Eq. 1, it can be considered k is a threshold value related to packet loss $(0 < k < 1)$ as C is the finite cache memory size and $\alpha$ is the Zip f-parameter.

**Video data searching mechanism:** Let us consider $C_{(s)}$ is the equivalent capacity from the network with respect to bit streaming rate. Those were selected to ensure the aggregate stationary bit rate $B_{(s)}$ at any variable session (s). If the aggregate stationary bit rate exceeds $C_{(s)}$ i.e., $\{B_{(s)} \succ C_{(s)}\}$ then current request will be lost or simultaneously the requested portion of the chunk video data be discarded from the system. If $\{B_{(s)} \succ C_{(s)}\}$ for any session (s), then it was clearly indicate unused of the complete bandwidth. Eventually, the system provides fewer throughputs. So for the uniform data streaming for any session (s) the condition $\{B_{(s)} \leq C_{(s)}\}$ holds true. The Video on demand system requires atleast k' number of database server to be active in data streaming, out of the total number of data storage servers in any session (s). To maintain the delay and jitter at optimize levels, the hop count number k' should be within a limit without imposing any impact on the stable data streaming condition at the session (s). The aggregate stationary bit rate was required mostly for the approximation. The adaptive encoding runs at the video storage, to maintain a certain standard. The packet loss or bit loss probability, maintains the aggregate stationary bit rate which remains same value for a short burst period of time. So, $p(B_{(s)} \geq C_{(s)})$ was the packet loss or usually bit loss probability. The distribution of $B_{(s)}$ is generally obtained from the two-state Markov chain. During the data streaming, one individual active channel was required for the case of broadcasting or multicasting[12]. The enthusiastic link remains enthusiastic throughout the data transmission in the session (s) between the web cache servers via the application server to the data-storage. Here, $p_k$ presents the probability that k out of N video storage servers are enthusiastic. Enthusiastic means that the storage server has the required encoded video file and the link was actively live throughout the streaming at the session (s). So get the Eq. 6:

$$p_k \approx \binom{N}{k} \rho^k (1-\rho)^{N-k} \tag{6}$$

The binary event space is {0, 1} with the equal probability {½, ½} for $\rho$. The value of $C_{(s)}$ is obtained by computing the possible smallest integer k' (say). So, there should be at least k' number of database server that has the required chunk video object.

We have considered L and M are two events such that $M = (B_{(s)} \leq C_{(s)})$. Here, L = {At least k' number of database servers that can stream data out of the N database server at a session (s) from the storage, according to Fig. 2}.

The corresponding conditional dependency is presented with the following expression:

$$p\left(\frac{L}{M}\right) = \begin{bmatrix} \dfrac{p\left(\dfrac{M}{L}\right)p(L)}{p(M)} \Rightarrow p\left(\dfrac{L}{M}\right) \\ \dbinom{N}{k'+1}\rho^{k'+1}(1-\rho)^{N-k'-1} + \cdots + \cdots + \\ \dbinom{N}{k'+r}\rho^{k'+r}(1-\rho)^{N-k'-r} + \cdots \end{bmatrix} \tag{7}$$

for $r \in I^{-0}$.

According to the Eq. 7 clearly, $k' \leq N$ select the minimum number of database server out of from the N number of database server in storage domain. It was mandatory requirement, that there should be at least k' the number of database server those possessed the matched query string instigated from the viewer request. The query string for the video file was generated and broadcast from the application server inside the storage, according to stated algorithm "session based video search". The optimization problem becomes straight forward as to find the least integer value of k'. The following section presents the video searching procedure to identify the database server from where the video has to be retrieved. The database servers are distributed inside the storage domain. The searching procedure begins with the selection of application server. The application server generates the require query as the collection of strings. The strings are the headers of the buffer address and storage location of the video. This occurs due to case of hit misses at the web cache server. According to Eq. 5 the video data was retrieved from the distributed database inside the storage. In the case of new video stream, query was generated for the full stream video. In the case of commercial video on demand system, the supplier maintains their own network infrastructure and address. This was maintained by the private address system for security and billing purpose only.

---

**Algorithm:** "Session based video search"

```
// t is the time variable
// t_s is the system generated start time
// Δ is the session duration
// S is the collection of database nodes
INPUT: t, S, Δ
INITIALIZE:
Boolean: matched_node_not_found ← True
Q ← finite size queue
                        {S} ← set of data storage nodes
at t = t_s
        While (t<(t_s+Δ)) do
        Q ← Push {S} // first in first out
                While (match_node_not_found) do
                {
                v ← Pop from Q
                                        For each {w} from single hop neighbor of v

                        {
                        If (w possess the query string)
                        then
                        {
                                                match_node_not_found ← false
                                                OUTPUT: node ID of w
                                                    OUTPUT: data stream to query
                                                        initiative application server

                Exit
                        }
                                else
                        Q ← Push (w)
                        } // End for
        } //End While
t = t_s+Δ
 } // End While

OUTPUT: query matched node not found
End
```

---

## RESULTS AND DISCUSSION

Three stages of simulation have been conducted according to the aforesaid methodology. In the first stage of simulation, conducted traffic analysis determines the VOD system for inbound and outbound channel bandwidths that was required in a case of 'no blocking' service.

Golrezaei *et al.*[4] consider distributed storing of video and popularity distributions for the campus network. Golrezaei *et al.*[4], have considered an empirical distribution of file popularity from a wired network, namely the campus network of the University of Massachusetts at Amherst. They have considered an average number of active clusters versus deterministic caching. Their results show that the result depends on the parameter of the Zip f distribution, in particular the cases $\alpha > 1$ and $\alpha < 1$ leads to different results. In the special case $\alpha = 1$, they have omitted for simplicity. Golrezaei *et al.*[4], have noticed for $\alpha > 1$, the device to device video traffic throughput increases linearly with the number of users in the cluster.

To unraveling the impact of temporal and geographical locality in content caching systems, Traverso *et al.*[28], have evaluated the volume of user requests over three weeks for video traffic trace data. The cumulative number of requests over the time for a subset of storage videos, the cache size depends on the (evaluated number of distinct videos) presents and that the required to achieve. The cache presented a desired hit probability, when an LRU cache was fed by the requests contained.

Poularakis *et al.*[29] have explored the caching and multicast for 5G wireless networks. Poularakis *et al.*[29], obtained energy cost to achieve the multicast time period and small cache size in each of the small cell base stations (SBS).

Zhan and Wen[30] have considered caching ability to alleviate the pressure brought by the explosion of mobile video traffic on present architecture of cellular networks. Zhan and Wen[30] have considered small cell base stations (SBS) with limited caching ability to reduce the load at small cell base stations.

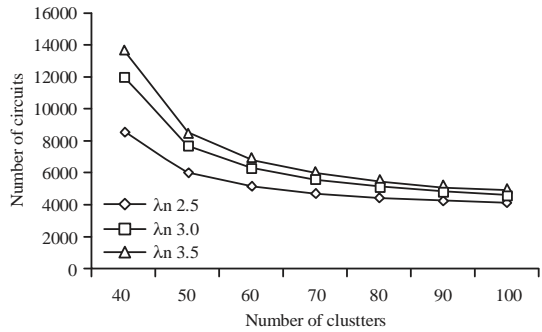

Fig. 3: Clustering versus bandwidth

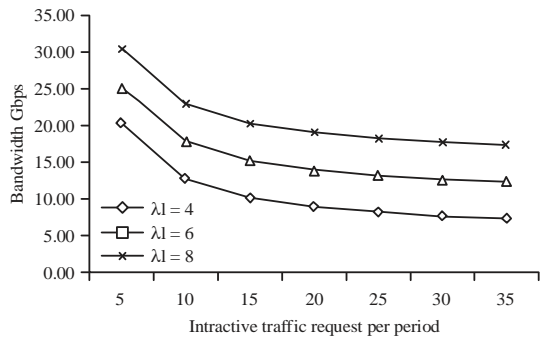

Fig. 4: Bandwidth requirements at interactive session

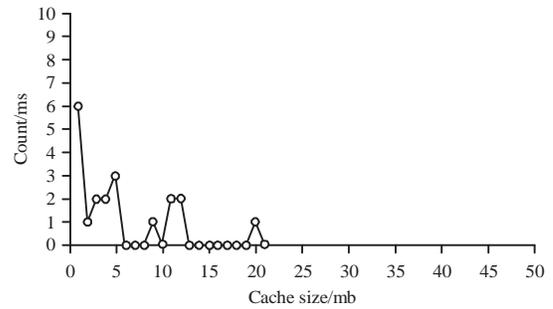

Fig. 5: Cache size 20, number of the submitted requests is 20 in 1 ms

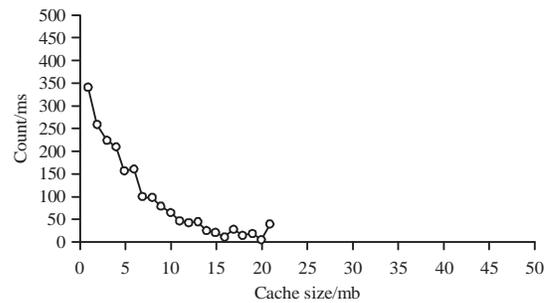

Fig. 6: Cache size 20, number of the submitted requests is 2000 in 1 ms

Saha *et al.*[31] have used the cache retention ratio with variable data population for the efficient cache availability management in information-centric networks.

Figure 3 presents the bandwidth requirement for the number of viewer clusters with versatile viewer's requests types and varying rates from 2.5, 3.0 and 3.5 Gbps, respectively.

The Fig. 4 presents bandwidth requirement in the case of an interactive session for the generated viewer requests, with 4, 6 and 8 Gbps rates during the unit session period. It is clear from the Fig. 3 that the VOD system with the help of clustering significantly reduces the required bandwidth. The Fig. 4 shows the required load come downs for any types of the interactive session likes pause/skip/hold/fast-forwards, when the number of cluster was increases.

In the second stage of simulation, it has been considered distributed database storages that were well connected to the application servers. The simulation results were obtained according to the proposed LRFU mechanism. The LRFU mechanisms are implemented at the web cache server with small size of cache memory are presented in Fig. 5 and 6.

Figure 5 presents the LRFU performance for the considered cache size 20. The number of submitted requests processed at the web cache memory was 20 in 1 ms. Figure 6 presents the LRFU performance for the considered cache size 20. The number of submitted requests processed at the web cache memory is 2000 in 1 ms. In the third stage of simulation, it has been considered a number of phases to make complete simulations with the different number of application server and database servers. First, it has been considered the number of application server is 14 and each application server was connected with a single database server. In the next phase of simulation, it has been considered eight application servers are there and each application server was connected with two database servers. The video data was placed at the remote database storage server. In the advanced stage of simulation, it has been considered seven application servers and every application server was well connected with three data-base storage servers. On the same way for the advanced phase of simulation, it has been considered six application servers. Those are well connected and each application server was connected with five data-storage servers. In every phase of simulation it has been considered one minute as a session interval. The total simulation was run for 10 min. The simulation graphs have 2 axis. Horizontal axis presents the hop counts towards the searching database node.

The vertical axis presents the number of request generated at application server and that successfully received video stream from the remote database nodes. Video data were transmitted between the database nodes and application servers through reliable TCP protocol. Figure 7 presents data flow graph that was more or less similar to Gaussian distribution. The major requests were served by database node with the different hop counts that was reflected in the left branch of the curve. The number of hop counts in the left branch is 7 and at the right branch it is 6, It has been observed that the maximum request served by the hop number 7. At the next stage, 8 application servers being considered and each application server were connected with two data-storage servers. Here, the major requests were served from the nearer database nodes. It is presented in the Fig. 8. The major requests were served from the database node with hop count 5. In final stages of simulation, it has been considered 6 and 7 application servers, respectively.

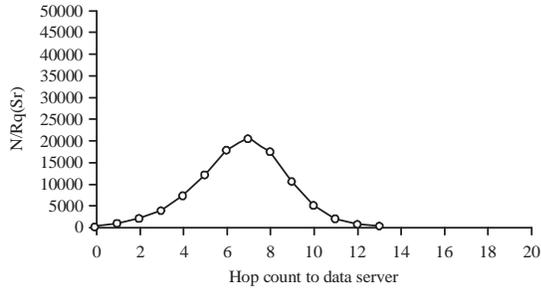

Fig. 7: Storage contains one database server

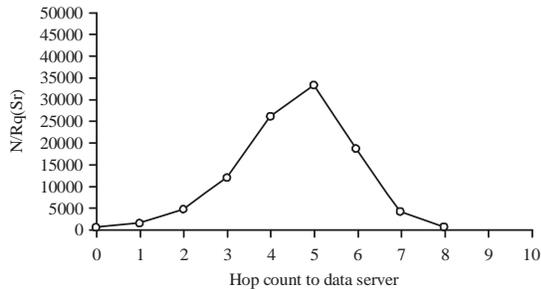

Fig. 8: Storage contains two database servers

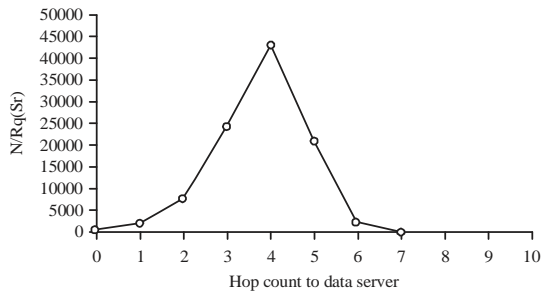

Fig. 9: Storage contains three database servers

Three database nodes were connected with one application server. In very next phase, the simulation was considered 5 database nodes those were connected with the single application server. The final phase of simulation presents in Fig. 9. Figure 9 presents the major request served from the nearest database storage servers. It has been shown the hop counts gradually decrease even if the application server decreases. If the application server was decreasing, then the system maintaining cost was decreasing and the searching cost also decreases. The jitter and packet delay was directly depended on the number of hop counts. In general scenario, if the hop's count for searching was increased then the impact was observed in the packet jitter. The multi path packet forwarding algorithm was not efficient for the data transfer from the database node to the application server. In this scenario, the multi path data stream forward procedure could increase unnecessary packet delay. It leads to extra metric with addition to the data transportation problem. If the link was closed during the data transmission from the database node to the application server, then it was useful to use multipath adaptive routing algorithm. In case of huge content delivery from the database node to the application server, it was always preferable to follow the dedicated established link, which reduces the delay for stream transfer. From the following three graphs, i.e., Fig. 7, 8 and 9, observed that 20000 requests are transmitted from hop count 7 in Fig. 7. About 30000 requests are transmitted from the hop count 5 in Fig. 8 and 42000 numbers of requests are transmitted from hop count 4 in Fig. 9.

It implicates, a balance distribution of video data between the database's nodes to the application server, usually reduces the searching cost. It effectively reduces the delay and packet drop. A simple estimate to measure the performance of the proposed methodology, a metric was considered as the sum of the 'number of request serve' $c_i$ multiply with the hop counts $s_i$ i.e., $\sum_{i}^{m}(c_i \times s_i)$. The expression gave the following performance score. The score value according to Fig. 7 is 662000. The score value according to Fig. 8 is 460000. The score value according to the Fig. 9 is 368000. The score value was equivalent to the number of computation require to transmit video chunk data form the database node to web cache through the application servers. Clearly, the performance of the system enhanced as the score value decreases.

## CONCLUSION

The study presents efficient traffic analysis in video on demand system to minimize the hop counts to explore the exact video storage node. It is presented the requirement of traffic control at the web cache server to enhance the traffic flow in the video on demand system. This studies address the requirement to design and develops an efficient robust back bone architecture based on multi-tier distributed hybrid architecture. This is to be integrated with distributed data storage. The smooth transportation of video data is highly required for versatile areas of deployment. To enhance the performance of video on demand system, traffic analysis and traffic control is required at various levels inside the VOD system. The major areas of deployments can be categorized in, academic, entertainment, surveillance, counter terrorism. The session based video searching procedure successfully reduces the hop counts, from the web cache server to the database nodes. This study has shown that in the high-speed network, video data stored among the different distributed nodes, which enhanced the system performance, instead to store in a single storage node.

## SIGNIFICANCE STATEMENTS

This study discovers the effectiveness of traffic analysis to explore the exact media storage in video on demand system with limited hop counts. This study is beneficial for the researcher to uncover the critical areas of traffic analytic, that many researchers were not able to explore. Thus a new methodology emerges about the session oriented aggregate bit streaming rate combined with the total capacity of the VOD (video on demand) network, provides important prior knowledge about the current load inside the VOD system.

## REFERENCES

1. Cisco Visual Networking Index, 2017. Global mobile data traffic forecast update 2016-2021. White Paper, Document ID: 1454457600805266, Cisco Visual Networking Index.
2. Kanrar, S., 2012. Analysis and implementation of the large scale video-on-demand system. Int. J. Applied Inform. Syst., 1: 41-49.
3. Hsu, W.W., A.J. Smith and H.C. Young, 2005. The automatic improvement of locality in storage systems. ACM Trans. Comput. Syst., 23: 424-473.
4. Golrezaei, N., A.F. Molisch, A.G. Dimakis and G. Caire, 2013. Femtocaching and device-to-device collaboration: A new architecture for wireless video distribution. IEEE. Commun. Mag., 51: 142-149.
5. Huang, C., J. Li and K.W. Ross, 2007. Can internet video-on-demand be profitable? Proceedings of the Conference on Applications, Technologies, Architectures and Protocols for Computer Communications, August 27-31, 2007, Kyoto, Japan, pp: 133-144.
6. Vilas, M., X.G. Paneda, R. Garcia, D. Melendi and V.G. Garcia, 2005. User behavior analysis of a video-on-demand service with a wide variety of subjects and lengths. Proceedings of the 31st EUROMICRO Conference on Software Engineering and Advanced Applications, August 30-September 3, 2005, Porto, Portugal, pp: 330-337.
7. Cha, M., H. Kwak, P. Rodriguez, Y.Y. Ahn and S. Moon, 2009. Analyzing the video popularity characteristics of large-scale user generated content systems. IEEE/ACM Trans. Network., 17: 1357-1370.
8. Gill, P., M. Arlitt, Z. Li and A. Mahanti, 2007. Youtube traffic characterization: A view from the edge. Proceedings of the 7th ACM SIGCOMM Conference on Internet Measurement, October 24-26, 2007, San Diego, California, USA., pp: 15-28.
9. Kanrar, S., 2011. Performance of distributed video on demand system for multirate traffic. Proceedings of the IEEE Conference on Recent Trends in Information Systems, December 21-23, 2011, Kolkata, India, pp: 52-56.
10. Tian, Y., H.M.S. Edwin, C. Chantrapornchai and P.M. Kogge, 1998. Optimizing page replacement for multiple-level memory hierarchy. https://pdfs.semanticscholar.org/d6d9/6a5945db34d568c5be88bca2422968ba842e.pdf
11. Lee, W., S. Park, B. Sung and C. Park, 2011. Improving adaptive replacement cache (ARC) by reuse distance. Proceedings of the 9th USENIX Conference on File and Storage Technologies, February 15-17, 2011, San Jose, CA, USA.
12. Muntz, D. and P. Honeyman, 1992. Multi-level caching in distributed file systems-or-your cache a in't nuthin' but trash. Proceedings of the Usenix Winter 1991 Conference, January 1991, Dallas, TX., USA.
13. Ari, I., A. Amer, R.B. Gramacy, E.L. Miller, S.A. Brandt and D.D.E. Long, 2002. ACME: Adaptive caching using multiple experts. Proceedings of the 4th International Workshop on Distributed Data and Structures, March 20-23, 2002, Paris, France, pp: 143-158.
14. Zhou, Y., J.F. Philbin and K. Li, 2001. The multi-queue replacement algorithm for second level buffer caches. Proceedings of the USENIX Annual Technical Conference, June 25-30, 2001, Boston, MA., USA., pp: 91-104.
15. Kanrar, S. and N.K. Mandal, 2017. Video traffic analytics for large scale surveillance. Multimedia Tools Applic., 76: 13315-13342.
16. Liu, Y., Y. Guo and C. Liang, 2008. A survey on peer-to-peer video streaming systems. Peer-to-peer Network. Applic., 1: 18-28.
17. Almeida, V., A. Bestavros, M. Crovella and A. de Oliveira, 1996. Characterizing reference locality in the WWW. Proceedings of the 4th International Conference on Parallel and Distributed Information Systems, December 18-20, 1996, Miami Beach, FL, USA., pp: 92-103.

18. Kanrar, S., N.K. Mandal and S.D. Kanrar, 2015. Session based storage finding in video on demand system. Proceedings of the 3rd International Symposium on Women in Computing and Informatics, August 10-13, 2015, Kochi, India, pp: 131-135.

19. Guerin, R., H. Ahmadi and M. Naghshineh, 1991. Equivalent capacity and its application to bandwidth allocation in high-speed networks. IEEE J. Selected Areas Commun., 9: 968-981.

20. Balamash, A. and M. Krunz, 2004. An overview of web caching replacement algorithms. IEEE Commun. Surveys Tutorials, 6: 44-56.

21. Kanrar, S., 2011. Efficient traffic control of VOD system. Int. J. Comput. Networks Commun., 3: 95-106.

22. Kanrar, S. and K.N. Mandal, 2016. Video traffic flow analysis in distributed system during interactive session. Adv. Multimedia, Vol. 2016. 10.1155/2016/7829570.

23. Kanrar, S. and K.N. Mandal, 2015. Efficient video streaming for interactive session. Proceedings of the 8th Annual ACM India Conference, October 29-31, 2015, Ghaziabad, India, pp: 79-83.

24. Abrams, M., C.R. Standridge, G. Abdulla, E.A. Fox and S. Williams, 1996. Removal policies in network caches for world-wide web documents. Proceedings of the Conference on Applications, Technologies, Architectures and Protocols for Computer Communications, August 28-30, 1996, Palo Alto, California, USA., pp: 293-305.

25. Shah, M., O. Javed and K. Shafique, 2007. Automated visual surveillance in realistic scenarios. IEEE Multimedia, 14: 30-39.

26. Kanrar, S. and K.N. Mandal, 2014. Dynamic Page Replacement at the Cache Memory for the Video on Demand Server. In: Advanced Computing, Networking and Informatics-Volume 2, Kundu, M.K., D.P. Mohapatra, A. Konar and A. Chakraborty (Eds.). Springer, Cham, ISBN: 978-3-319-07349-1, pp: 461-469.

27. Lee, Y., I. Park and Y. Choi, 2002. Improving TCP performance in multipath packet forwarding networks. J. Commun. Networks, 4: 148-157.

28. Traverso, S., M. Ahmed, M. Garetto, P. Giaccone, E. Leonardi and S. Niccolini, 2015. Unravelling the impact of temporal and geographical locality in content caching systems. IEEE Trans. Multimedia, 17: 1839-1854.

29. Poularakis, K., G. Iosifidis, V. Sourlas and L. Tassiulas, 2016. Exploiting caching and multicast for 5G wireless networks. IEEE Trans. Wireless Commun., 15: 2995-3007.

30. Zhan, C. and Z. Wen, 2017. Content cache placement for scalable video in heterogeneous wireless network. IEEE Commun. Lett., Vol. PP, Issue 99. 10.1109/LCOMM.2017.2756033.

31. Saha, S., A. Lukyanenko and A. Yla-Jaaski, 2015. Efficient cache availability management in information-centric networks. Comput. Networks, 84: 32-45.